\documentclass[11pt]{article} 

\usepackage[utf8]{inputenc} 
\usepackage{algorithm,algorithmic,natbib}

\usepackage{geometry} 
\usepackage{graphicx} 

\usepackage{booktabs} 
\usepackage{array} 
\usepackage{paralist} 
\usepackage{verbatim} 
\usepackage{subfig} 

\usepackage{fancyhdr} 
\usepackage{sectsty}
\allsectionsfont{\sffamily\mdseries\upshape} 

\usepackage[nottoc,notlof,notlot]{tocbibind} 
\usepackage[titles,subfigure]{tocloft} 

\begin{document}
\vspace*{1in}

\begin{center}
{\Large\bf Augmentation Schemes for Particle MCMC} 

\vspace{.8cm}

Paul Fearnhead$^{1*}$, Loukia Meligkotsidou$^{2}$. 
\end{center}

\vspace{.3cm}

\textit{1. Department of Mathematics and Statistics, Lancaster University. \\
2. Department of Mathematics, University of Athens.\\
}

\textit{
* Correspondence should be addressed to Paul Fearnhead. (e-mail:
p.fearnhead@lancaster.ac.uk).}

\vspace{.4cm}

\parindent=0mm

\begin{center}
\textbf{Abstract}
\end{center}
Particle MCMC involves using a particle filter within an MCMC algorithm. For inference of a model which
involves an unobserved stochastic process, the standard implementation
uses the particle filter to propose new values for the stochastic process, and MCMC moves to propose new values
for the parameters. We show how particle MCMC can be generalised beyond this. Our key idea is to introduce
new latent variables. We then use the MCMC moves to update the latent variables, and the particle filter to 
propose new values for the parameters and stochastic process given the latent variables. A generic way of defining these latent
variables is to model them as pseudo-observations of the parameters or of the stochastic process. By choosing the amount
of information these latent variables have about the parameters and the stochastic process we can often improve
the mixing of the particle MCMC algorithm by trading off the Monte Carlo error of the particle filter and the mixing
of the MCMC moves. We show that using pseudo-observations within particle MCMC can improve its efficiency in
certain scenarios: dealing with initialisation problems of the particle filter; speeding up the mixing of particle Gibbs when there
is strong dependence between the parameters and the stochastic process; and 
enabling further MCMC steps to be used within the particle filter.


\section{Introduction}

Particle MCMC \cite[]{Andrieu/Doucet/Holenstein:2010} is a recent extension of MCMC. It is most naturally applied to inference for models, such as state-space models, where there is an unobserved stochastic process.
Standard MCMC algorithms, such as Gibbs samplers, can often struggle with such models due to strong dependence between the unobserved process and the parameters \cite[see e.g.][]{Pitt/Shephard:1999b,Fearnhead:2011}.
Alternative Monte Carlo methods, called particle filters, can be more efficient for inference about the unobserved process given known parameter values, but struggle when dealing with unknown parameters. The idea of particle
MCMC is to embed a particle filter within an MCMC algorithm. The particle filter will then update the unobserved process given a specific value for the parameters, and MCMC moves will be used to update the parameter values.
Particle MCMC has already been applied widely: in areas such as econometrics \cite[]{Pitt:2012}, inference for epidemics \cite[]{Rasmussen:2011}, systems biology \cite[]{Golightly/Wilkinson:2011}, and probabilistic programming \cite[]{Wood:2014}.

The standard implementation of particle MCMC is to use an MCMC move to update the parameters and a particle filter to update the unobserved stochastic process \cite[though see][for alternatives]{Murray:2012,Wood:2014}.
However, this may be inefficient, due to a large Monte Carlo error in the particle filter, or due to slow mixing of the MCMC moves. The idea of this paper is to consider generalisations of this standard implementation, which can
lead to more efficient particle MCMC algorithms.

In particular, we suggest a data augmentation approach, where we introduce new latent variables into the model. We then implement particle MCMC on the joint 
posterior distribution of the parameters, unobserved stochastic process and latent variables. We use MCMC to update the latent variables and a particle filter to update
the parameters and the stochastic process. The intuition behind this approach is that the latent variables can be viewed as containing information about the parameters and the stochastic process.
The more information they contain, the lower the Monte Carlo error of the particle filter. However, the more information they contain, the stronger the dependencies in the posterior distribution, and hence the
poorer the MCMC moves will mix. Thus, by carefully choosing our latent variables, we are able to appropriately trade-off the error in the particle filter against the mixing of the MCMC, so as to improve the 
efficiency of the particle MCMC algorithm.

In the next section we introduce particle filters and particle MCMC. Then in Section \ref{S:EAS} we introduce our data augmentation approach. A key part of this is constructing a generic way of defining the latent variables
so that the resulting particle MCMC algorithm is easy to implement. This we do by defining the latent variables to be observations of the parameters or of the stochastic process. By defining the likelihood for
these observations to be conjugate to the prior for the parameters or the stochastic process we are able to analytically calculate quantities needed to implement the resultant particle MCMC algorithm. Furthermore, the accuracy
of the pseudo-observations can be varied to allow them to contain more or less information. In Section \ref{S:EX} we investigate the efficiency of the new particle MCMC algorithms. We focus on three scenarios where we
believe the data augmentation approach may be particularly useful. These are to improve the mixing of the particle Gibbs algorithm when there are strong dependencies between parameters and the unobserved stochastic process; to 
enable MCMC to be used within the particle filter; and to deal with diffuse initial distributions for the stochastic process. The paper ends with a discussion.

\section{Particle MCMC}


\subsection{State-Space Models}

For concreteness we consider application of particle MCMC to a state-space model, though both particle MCMC and the ideas we develop in this paper can be applied more generally. Throughout we will use $p(\cdot)$ and $p(\cdot|\cdot)$  to denote general marginal and conditional probabililty density functions, with the arguments making it clear which distributions these relate to.

Our state-space model will be parameterised by $\theta$, and we introduce a prior distribution for this parameter, $p(\theta)$.  We then have a latent discrete-time stochastic process, $X_{1:T}=(X_1,\ldots,X_T)$. This process is often termed the state-process, and we assume it is a Markov process. Thus the probability density of a realisation of the state process can be written as
\[
p(x_{1:T}|\theta)=p(x_1|\theta)\prod_{t=2}^T p(x_t|x_{t-1},\theta).
\]

We do not observed the state process directly.  Instead we take partial observations at each time-point, $y_{1:T}=(y_1,\ldots,y_T)$. 
We assume that the observation at any $t$ just depends on the state process through its value at that time, $x_t$. Thus we can write the likelihood of the observations, given the state process and parameters as,
\[
p(y_{1:T}|x_{1:T},\theta)=\prod_{t=1}^T p(y_t|x_t,\theta).
\]

Our interest is in calculating, or approximating, the posterior for the parameters and states:
\begin{eqnarray}
p(x_{1:T},\theta|y_{1:T}) &\propto& p(\theta) p(x_{1:T}|\theta) p(y_{1:T}|x_{1:T},\theta) \nonumber \\
&=& p(\theta)\left[ p(x_1|\theta)\prod_{t=2}^T p(x_t|x_{t-1},\theta)\right]\left[ \prod_{t=1}^T p(y_t|x_t,\theta)\right]. \label{eq:1}
\end{eqnarray}

We frequently use the notation of an extended state vector $\mathcal{X}_t=(x_{1:t},\theta)$, which consists of the full path of the state process to time $t$, and the value of the parameter. 
Thus $\mathcal{X}_T$ consists of the full state-process and the parameter, and we are interested in calculating or approximating $p(\mathcal{X}_T|y_{1:T})$.

\subsection{Particle Filters}

Particle filters are Monte Carlo algorithms that can be used to approximate posterior distributions for state-space models, such as (\ref{eq:1}). We will describe a particle filter with a view to their use within particle MCMC algorithms, which is introduced in the next section. 

Rather than use a particle filter to approximate (\ref{eq:1}), we will consider conditioning on part of the process. We will assume we condition on $\mathcal{Z}$, which will be some function of $\mathcal{X}_T$.  Thus the particle filter will target $p(\mathcal{X}_T|\mathcal{Z},y_{1:T})=p(x_{1:T},\theta|\mathcal{Z},y_{1:T})$. The output of running the particle filter will just be an estimate of the marginal likelihood for the data given $\mathcal{Z}$, and a single realisation for the parameter and state process. A simple particle filter algorithm is given in Algorithm \ref{Alg:1}. 

\begin{algorithm}
   \caption{Particle Filter Algorithm}
   \label{Alg:1}
{\bf Input:} \\
A value of $\mathcal{Z}$. \\
The number of particle, $N$.
   \begin{algorithmic}[1]
     \FOR{$i=1,\ldots,N$}  
      \STATE Sample $\mathcal{X}_1^{(i)}$ independently from $p(\mathcal{X}_1|\mathcal{Z})$.
     \STATE Calculate weights $w_1^{(i)}=p(y_1|\mathcal{X}_1^{(i)},\mathcal{Z})$.
    \ENDFOR
     \STATE Set $\hat{p}(y_1|\mathcal{Z})=\frac{1}{N} \sum_{i=1}^Nw_1^{(i)}$.
      \FOR{$t=2,\ldots,T$}
      \FOR{$i=1,\ldots,N$}
         \STATE Sample $j$ from $\{1,\ldots,N\}$ with probabilities proportional to $\{w_{t-1}^{(1)},\ldots,w_{t-1}^{(N)}\}$. 
        \STATE Sample $\mathcal{X}_t^{(i)}$ from $p(\mathcal{X}_t|\mathcal{X}_{t-1}^{(j)},\mathcal{Z})$.
         \STATE Calculate weights $w_t^{(i)}=p(y_t|\mathcal{X}_t^{(i)},\mathcal{Z})$.
	\ENDFOR
         \STATE Set $\hat{p}(y_{1:t}|\mathcal{Z})=\hat{p}(y_{1:t-1}|\mathcal{Z})\left(\frac{1}{N} \sum_{i=1}^N w_t^{(i)}\right)$.
      \ENDFOR
\STATE Sample $j$ from $\{1,\ldots,N\}$ with probabilities proportional to $\{w_{T}^{(1)},\ldots,w_{T}^{(N)}\}$
   \end{algorithmic}
{\bf Output:} A value of the extended state, $\mathcal{X}_T^{(j)}$, and an estimate of the marginal likelihood $\hat{p}(y_{1:T}|\mathcal{Z})$. 
\end{algorithm}

If we stopped this particle filter algorithm at the end of iteration $t$, we would have a set of values for the extended state, often called particles, each with an associated weight. 
These weighted particles give an approximation to $p(\mathcal{X}_t|y_{1:t},\mathcal{Z})$. At iteration $t+1$ we propagate the particles and use importance sampling to create a set of weighted particles 
to approximate $p(\mathcal{X}_{t+1}|y_{1:t+1},\mathcal{Z})$. This involves first generating new particles at time $t+1$ through (i) sampling particles from the approximation to $p(\mathcal{X}_t|y_{1:t},\mathcal{Z})$; and (ii)
propagating these particles by simulating values for $X_{t+1}$ from the transition density of the state-process, $p(X_{t+1}|\mathcal{X}_t)$. Secondly, each of these particles at $t+1$ is then given a weight
proportional to the likelihood of the observation $y_{t+1}$ for that particle value. \cite[See][for more details]{Doucet/Godsill/Andrieu:2000,Fearnhead:2008}. 
At the end of iteration $T$ we output a single value of $\mathcal{X}_{T}$, by sampling once from the particles at time $T$, with the probability of choosing a particle being proportional to its weight.
 
A by product of the importance sampling at iteration $t+1$ is that we get a Monte Carlo estimate of $p(y_{t+1}|y_{1:t},\mathcal{Z})$, 
and the product of these for $t=1,\ldots,T$ gives an unbiased estimate of the marginal likelihood $p(y_{1:T}|\mathcal{Z})$ \cite[][proposition 7.4.1]{DelMoral:2004}. 
This unbiased estimate will be key to the implementation of particle MCMC, and is also output.

This is arguably the simplest particle filter implementation, and more efficient extensions do exist \cite[see, for example][]{Liu/Chen:1998,Pitt/Shephard:1999,Carpenter/Clifford/Fearnhead:1999,Gilks/Berzuini:2001}. 
In describing this algorithm we have implicitly assumed that $\mathcal{Z}$ has been chosen so that all the conditional distributions needed for implementing particle filter are easy to calculate or sample from, 
as required. In practice this will mean $\mathcal{Z}$ depending on either the parameters and/or the initial state or states of the latent process. 
The most common choice would be to condition on the parameter values, $\mathcal{Z}=\theta$, or to have no conditioning, $\mathcal{Z}=\emptyset$. 
The latter choice means that parameter values are sampled within the particle filter, and can often lead to particle degeneracy: 
whereby most or all particles at time $T$ have the same parameter value. This is less of an issue when we are implementing particle MCMC, as we output a single particle, 
than when using particle filters to approximate the posterior distribution \cite[][Section 2.2.2]{Andrieu/Doucet/Holenstein:2010}.

\subsection{Particle MCMC} \label{S:PMCMC}

The idea of particle MCMC is to use a particle filter within an MCMC algorithm. There are two generic implementations of particle MCMC: particle marginal Metropolis-Hastings (PMMH) and particle Gibbs. 

\subsubsection{Particle marginal Metropolis-Hastings Algorithm}

First we describe the particle marginal Metropolis-Hastings (PMMH) sampler \cite[][Section 2.4.2]{Andrieu/Doucet/Holenstein:2010}.  This involves choosing, $\mathcal{Z}$, an appropriate function of 
the extended state $\mathcal{X}_T$. Our MCMC algorithm has a state that is $\{\mathcal{Z},\mathcal{X}_T,\hat{p}(y_{1:T}|\mathcal{Z})\}$, a value for this function, a corresponding value for the extended state and an 
estimate for the marginal likelihood given the current value of $\mathcal{Z}$. We assume that $\mathcal{Z}$ has been chosen so that we can both implement the particle filter of Algorithm \ref{Alg:1}, 
and also that we can calculate the marginal distribution, $p(\mathcal{Z})$. A common choice is $\mathcal{Z}=\theta$, though see below for other possibilities.

Within each iteration of PMMH we first propose a new value for $\mathcal{Z}$, using a random walk proposal. 
Then we run a particle filter to both propose a new value of $\mathcal{X}_T$ and to calculate an estimate for the marginal likelihood. These new values are then accepted with a probability that depends on the ratio of the new and old estimates of the marginal likelihood. 
Full details are given in Algorithm \ref{Alg:2}. 
\begin{algorithm}
   \caption{Particle marginal Metropolis-Hastings Algorithm \cite[]{Andrieu/Doucet/Holenstein:2010}}
   \label{Alg:2}
 {\bf Input:} \\
An initial value $\mathcal{Z}^{(0)}$. \\
A proposal distribution $q(\cdot|\cdot)$.\\
The number of particles, $N$, and the number of MCMC iterations, $M$.
  \begin{algorithmic}[1]
\STATE Run Algorithm \ref{Alg:1} with $N$ particles, conditioning on $\mathcal{Z}^{(0)}$, to obtain $\mathcal{X}_T^{(0)}$ and $\hat{p}(y_{1:T}|\mathcal{Z}^{(0)})$. 
     \FOR{$i=1,\ldots,M$}  
      \STATE Sample $\mathcal{Z}'$ from $q(\mathcal{Z}|\mathcal{Z}^{(i-1)})$.
     \STATE Run Algorithm \ref{Alg:1} with $N$ particles, conditioning on $\mathcal{Z}'$, to obtain $\mathcal{X}_T'$ and $\hat{p}(y_{1:T}|\mathcal{Z}')$. 
    \STATE With probability
\[
\min\left\{ 1, \frac{q(\mathcal{Z}^{(i-1)}|\mathcal{Z}')\hat{p}(y_{1:T}|\mathcal{Z}')p(\mathcal{Z}')}{q(\mathcal{Z}'|\mathcal{Z}^{(i-1)})\hat{p}(y_{1:T}|\mathcal{Z}^{(i-1)})p(\mathcal{Z}^{(i-1)})} \right\}
\]
set $\mathcal{X}_T^{(i)}=\mathcal{X}_T'$, $\mathcal{Z}^{(i)}=\mathcal{Z}'$ and $\hat{p}(y_{1:T}|\mathcal{Z}^{(i)})=\hat{p}(y_{1:T}|\mathcal{Z}')$; otherwise set $\mathcal{X}_T^{(i)}=\mathcal{X}_T^{(i-1)}$,
$\mathcal{Z}^{(i)}=\mathcal{Z}^{(i-1)}$ and $\hat{p}(y_{1:T}|\mathcal{Z}^{(i)})=\hat{p}(y_{1:T}|\mathcal{Z}^{(i-1)})$
      \ENDFOR
\end{algorithmic}
{\bf Output:} A sample of extended state vectors: $\{\mathcal{X}_T^{(i)}\}_{i=1}^M$.
\end{algorithm}

One intuitive interpretation of this algorithm is that  we are using a particle filter to sample a new value of $\mathcal{X}'_T$ from an approximation to $p(\mathcal{X}'_T|\mathcal{Z}')$ within a standard MCMC algorithm. 
If we ignore the approximation, and denote the current state by $(\mathcal{X}_T,\mathcal{Z})$, then the acceptance probability of this MCMC algorithm would be
\[
\min\left\{1, \frac{q(\mathcal{Z}|\mathcal{Z}')p(y_{1:T}|\mathcal{Z}')p(\mathcal{Z}')}{q(\mathcal{Z}'|\mathcal{Z})p(y_{1:T}|\mathcal{Z})p(\mathcal{Z})} \right\},
\]
as the $p(\mathcal{X}'_T|\mathcal{Z}')$ terms cancel as they appear in both the target and the proposal.
The actual acceptance probability we use just replaces the, unknown, marginal likelihoods with our estimates. 
The magic of particle MCMC is that despite these two approximations, both in the proposal distribution for $\mathcal{X}_T$ given $\mathcal{Z}$ and in the marginal likelihoods, 
the resulting MCMC algorithm has the correct stationary distribution.

\subsubsection{Particle Gibbs}

The alternative particle MCMC algorithm, particle Gibbs, aims to approximate a Gibbs sampler. A Gibbs sampler that targets $p(\mathcal{Z},\mathcal{X}_T|y_{1:T})$ 
would involve iterating between (i) sampling a new value for $\mathcal{Z}$ from its full-conditional given the other components of $\mathcal{X}_T$; and (ii)
sampling a new value for $\mathcal{X}_T$ from its full conditional given $\mathcal{Z}$, $p(\mathcal{X}_{1:T}|\mathcal{Z},y_{1:T})$.

Implementing step (i) is normally straightforward. For example if $\mathcal{Z}=\theta$, this involves sampling new parameter values from their full conditional given the path of
the state-process. For many models, for example where there is conjugacy between the prior for the parameter and the model for the state and observation process, this distribution can be calculated analytically.

The difficulty, however, comes with implementing step (ii). The idea of particle Gibbs is to use a particle filter to approximate this step. Denote the current value for 
$\mathcal{X}_{1:T}$ by $\mathcal{X}^*_{1:T}$. Then, informally, this involves implementing a particle filter but conditioned on one of the particles at time $T$ being $\mathcal{X}^*_{1:T}$. We then sample one of the particles
at time $T$ from this conditioned particle filter and update $\mathcal{X}_{1:T}$ to the value of this particle. This simulation step is called a conditional particle filter, or a conditional SMC sampler. For full details
of this, and proof of the validity of the particle Gibbs sampler, see \cite{Andrieu/Doucet/Holenstein:2010}.

\subsubsection{Implementation}

Whilst both particle MCMC algorithms have the correct stationary distribution regardless of the accuracy of the particle filter, 
the accuracy does affect the mixing properties. More accurate estimates of the marginal likelihood will lead to more efficient algorithms \cite[]{Andrieu/Roberts:2009}. 
In implementing particle MCMC, as well as choosing details of the proposal distribution for $\mathcal{Z}$, we need also to choose the number of particles to use in the particle filter. 
Theory guiding these choices for PMMH is given in \cite{Pitt:2012}, \cite{Doucet:2012} and \cite{Sherlock:2013}. 

The standard implementation of particle MCMC will have $\mathcal{Z}=\theta$. 
However, our description is aimed to stress that particle MCMC is more general than this. It involves using MCMC proposals to update part of the extended state, 
and then a particle filter to update the rest. There is flexibility in choosing which part is updated by the MCMC move and which by the particle filter within the particle MCMC algorithm. 
For example, in order to deal with a diffuse initial distribution for the state-process, 
\cite{Murray:2012} choose $\mathcal{Z}=(\theta,X_1)$, so that MCMC is used to update both the parameters and the initial value of the state-process. Alternatively, \cite{Wood:2014} choose $\mathcal{Z}=\emptyset$, so that
both the parameters and the path of the state are updated using the particle filter.

To demonstrate this flexibility, and discuss its impact on the performance of the particle MCMC algorithm, we will consider a simple example.


\subsection{Example of Particle MCMC for linear-Gaussian Model}  \label{S:LGM}

We consider investigating the efficiency of particle MCMC for a simple linear-Gaussian model where we can calculate the posterior exactly. The model has a one-dimensional
state process, defined by 
\[
X_1=\sigma_1 \epsilon^{(X)}_1; ~~~~ X_t=\gamma X_{t-1}+\sigma_X \epsilon^{(X)}_t,\mbox{ for $t=2,\ldots,T$,}
\]
where $\epsilon^{(X)}_t$ are independent standard normal random variables. For $t=1,\ldots,T$ we have observations 
\[
 Y_t=\theta+X_t+\sigma_Y \epsilon^{(Y)}_t,
\]
where $\epsilon^{(Y)}_t$ are independent standard normal random variables. We assume that $\sigma_1^2$, $\sigma_X^2$, $\sigma_Y^2$ and $\gamma$ are known, and thus the only unknown parameter is $\theta$. Finally, we
assume a normal prior for $\theta$ with mean 0 and variance $\sigma_\theta^2$.

We simulated data for 100 times steps, with $\gamma=0.99$, $\sigma_Y=20$ and $\sigma_X$ chosen so that that $X_t$ process will have variance of 1 at stationarity. Our interest was in seeing how particle MCMC
performs in situations where there is substantial uncertainty in $X_1$ and $\theta$. Here we present results with $\sigma_\theta=100$ as we vary $\sigma_1$. We implemented particle MCMC with $\mathcal{Z}=\emptyset$,
$\mathcal{Z}=\{\theta\}$ and $\mathcal{Z}=\{\theta,X_1\}$. For the latter two implementations we used a random walk update for $\theta$ and $X_1$ with the variance set
to the posterior variance; with independent random walk updates for $\theta$ and $X_1$ when $\mathcal{Z}=\{\theta,X_1\}$.

\begin{figure}
\begin{center}
 \includegraphics[scale=0.5]{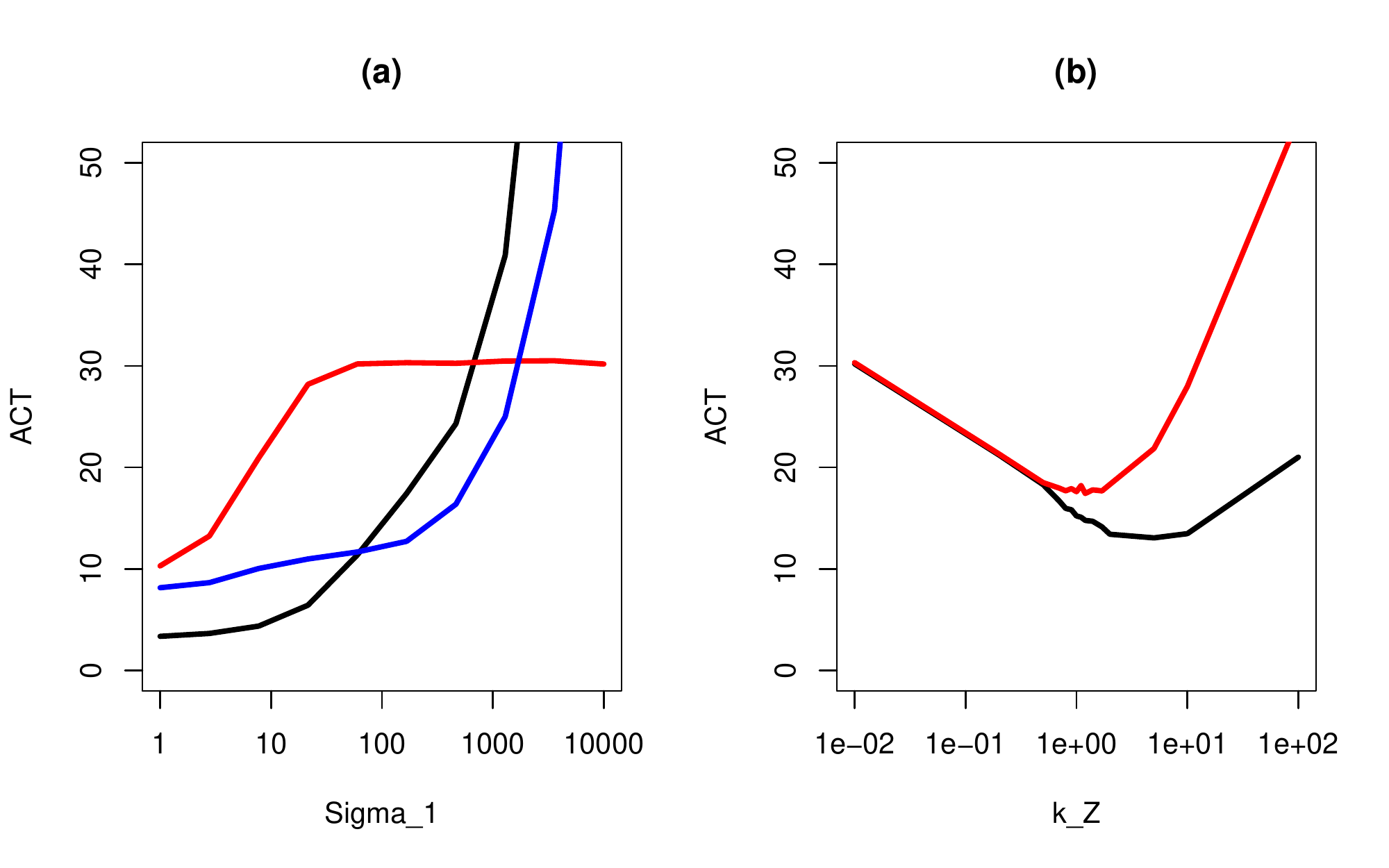}
 \caption{\label{Fig:LGSMM} Autocorrelation Times (ACT) for Particle MCMC runs of the Linear-Gaussian Model. (a) ACT for three particle MCMC algorithms as we vary $\sigma_1$: $\mathcal{Z}=\emptyset$ (black), 
 $\mathcal{Z}=\{\theta\}$ (blue), $\mathcal{Z}=\{\theta,X_1\}$ (red).  (b) ACT for particle MCMC with pseudo-observations, $\mathcal{Z}=(Z_x,Z_\theta)$ as we vary the variance of the noise in
 the definition of pseudo-observations; the variance of the noise for $Z_x$ and $Z_\theta$ is $k_Z$ times the marginal posterior variance for $X_1$ and $\theta$ respectively. ACT is shown for $X_1$ (black) and
 $Z_x$ (red).}
\end{center}
\end{figure}

To evalulate performance we ran each particle MCMC algorithm using 100 particles and $250,000$ iterations. We removed the first quarter of iterations as burn-in, and calculated autocorrelation times for estimating $\theta$. These 
are shown in Figure \ref{Fig:LGSMM}(a). 

The results show the trade-off in the choice of $\mathcal{Z}$. Including more information in $\mathcal{Z}$ leads to poorer mixing of the underlying MCMC algorithm, but comes at the advantage of smaller Monte Carlo
error in the estimate of the likelihood from the particle filter. This reduction in Monte Carlo error becomes increasingly important as the prior variance for $X_1$ increases. So for smaller values
of $\sigma_1$ the best algorithm has $\mathcal{Z}=\emptyset$, whereas when we increase $\sigma_1$ first the choice of $\mathcal{Z}=\{\theta\}$ then the choice of $\mathcal{Z}=\{\theta,X_1\}$ performs better. 

For this simple example there are alternative ways to improve the mixing of particle MCMC.
One reason that $\mathcal{Z}=\emptyset$ works better than $\mathcal{Z}=\{\theta,X_1\}$ for small $\sigma_1$ is that the posterior distribution has strong correlation between $X_1$ and $\theta$, 
and this correlation is ignored in the random walk update for $\mathcal{Z}=\{\theta,X_1\}$. So using a better reparameterisation \cite[]{Papaspiliopoulos/Roberts/Skold:2003}, or a better random walk proposal would
improve the mixing for the case where $\mathcal{Z}=\{\theta,X_1\}$. Also we have kept the number of particles fixed, whereas better results may be possible for $\mathcal{Z}=\emptyset$ and $\mathcal{Z}=\{\theta\}$ if we
increase the number of particles as we increase $\sigma_1$. However the main purpose of this example is to give insight into when different choices of $\mathcal{Z}$ would work well: adding information to
$\mathcal{Z}$ is one way to reduce the Monte Carlo error in estimates of the likelihood, but at the cost of slower mixing of the underlying MCMC algorithm particular if the posterior for $\mathcal{Z}$ has strong 
correlations which are not taken account of in the MCMC update.

\section{Augmentation Schemes for Particle MCMC} \label{S:EAS}

The example at the end of the previous section shows that the choice of which part of the extended state is updated by the particle filter, and which by a standard MCMC move, can have a sizeable 
impact on the performance of particle MCMC. Furthermore the default option for state-space models of updating parameters by MCMC and the state-process by a particle filter, is not always optimal. 

The potential within this choice can be greatly enhanced by augmenting the original model.
We will introduce an extra latent variable, $Z$, drawn from some distribution conditional on $\mathcal{X}_T$. This will introduce a new posterior distribution
\begin{equation} \label{eq:2}
p(\mathcal{X}_T,z|y_{1:T})=p(\mathcal{X}_T|y_{1:T})p(z|\mathcal{X}_T),
\end{equation}
where $p(\mathcal{X}_T|y_{1:T})$ is defined by (\ref{eq:1}) as before. 
 
For any choice of $p(z|\mathcal{X}_T)$, if we marginalise $Z$ out of (\ref{eq:2}) we get (\ref{eq:1}). Our approach will be to implement a particle MCMC algorithm for sampling from (\ref{eq:2}).  
This will give us samples $\{\mathcal{X}_T^{(i)},z^{(i)}\}_{i=1}^M$ from (\ref{eq:2}), with the $\{\mathcal{X}_T^{(i)}\}_{i=1}^M$ from (\ref{eq:1}) as required. 

In implementing the particle MCMC algorithm, using either PMMH 
or particle Gibbs, we will choose $\mathcal{Z}=z$. 
That is, we update the latent variable, $Z$, using the MCMC move, and we use a particle filter to update $\mathcal{X}_T$ conditional on $Z$. 
By appropriate choice of $p(z|\mathcal{X}_T)$ we hope to obtain a particle MCMC algorithm that mixes better than the standard implementation.

Whilst, in theory, we have a completely free choice over the distribution of the new latent variable, $Z$, in practice we need to be able to easily implement the resulting particle MCMC algorithm.
For both PMMH and particle Gibbs this will require us to be able to run a particle filter, or conditional particle filter, conditional on $Z$. In practice this will mean that we need to be able
to easily simulate from $p(\theta|z)$, $p(x_1|z,\theta)$ and, for $t=2,\ldots,T$, $p(x_t|\mathcal{X}_{t-1},z)$. For PMMH we will also need to be able to calculate the
acceptance probability of the algorithm involves, which involves the term
\begin{eqnarray*}
p(z)&=&\int p(z|\mathcal{X}_T)p(\mathcal{X}_T)\mbox{d}\mathcal{X}_T\\
&=& \int p(z|\theta,x_{1:T})p(\theta)p(x_1)\prod_{t=2}^T p(x_t|x_{t-1})\mbox{d}\theta\mbox{d}x_{1:T}.
\end{eqnarray*}
Thus we are restricted to cases where these conditional and marginal distributions can be calculated. We investigate possible generic choices in the next section.

\subsection{Generic Augmentation Schemes: Pseudo-Observations}

In choosing an appropriate latent variable $Z$ we need to first consider the ease with which we can implement the resulting particle MCMC algorithm.  
A generic approach is to model $Z$ as an observation of either $\theta$ or $x_1$ or both. As $Z$ is a latent variable we have added to the model, we call these pseudo-observations. 

By the Markov property of the state-process, if $Z$ only depends on $\theta$ and/or $x_1$ then we have $p(x_t|\mathcal{X}_{t-1},z)=p(x_t|\mathcal{X}_{t-1})$. Thus to be able to implement the particle filters we only
need to choose our model for the pseudo-observation so that we can simulate from $p(\theta|z)$ and $p(x_1|z,\theta)$. To enable this we can let each component of $Z$ be an independent pseudo-observation of a component
of $\theta$ or $x_1$, with the likelihood for the pseudo-observation chosen so that the prior for the relevant component of $\theta$ or $x_1$ is conjugate to this likelihood. Conjugacy will ensure that we can both simulate
from the necessary conditional distributions and we can calculate $p(z)$ as required to implement the particle MCMC algorithms. Constructing such models for the pseudo-observations is possible for many state-space models
of interest. In some applications other choices for $Z$ may be necessary or advisable: see Section \ref{S:DPMM} for an example.

To make these ideas concrete consider the linear Gaussian model of Section \ref{S:LGM}. We can choose $Z=(Z_\theta,Z_x)$ where $Z_\theta$ is a pseudo-observation of $\theta$ and $Z_x$ is one of $X_1$. As we have both
a Gaussian prior for $\theta$ and a Gaussian initial distribution for $X_1$, in each case a conjugate likelihood model arise from observations with additive Gaussian error. So for example we could choose
\begin{equation} \label{eq:ztheta}
 Z_\theta|\theta \sim \mbox{N}(\theta,\tau^2).
\end{equation}
This would give a marginal distribution of $Z_\theta\sim \mbox{N}(0,\tau^2+\sigma_\theta^2)$ and a conditional distribution of 
\[
 \theta|z_\theta \sim \mbox{N} \left(\frac{z_\theta \sigma_\theta^2}{\tau^2+\sigma_\theta^2},\frac{\tau^2\sigma_\theta^2}{\tau^2+\sigma_\theta^2} \right).
\]

Consider the case where we let $Z$ depend only on $\theta$. In specifying $p(z|\theta)$ we will have a choice as to how informative $Z$ is about $\theta$ -- for example the choice of $\tau$ in (\ref{eq:ztheta}) 
for the linear Gaussian model example. 
As such this gives a continuum between the implementations of particle MCMC in Section \ref{S:PMCMC}. In the limit as $Z$ is increasingly informative about $\theta$, 
we converge on an implementation of particle MCMC where we update $\theta$ using MCMC and $X_{1:T}$ using the particle filter. As $Z$ becomes less informative, we would tend to an implementation of 
particle MCMC where both $\theta$ and $X_{1:T}$ are updated through the particle filter. Where on this continuum is optimal will depend on a trade-off between Monte Carlo error in our estimate of 
the marginal likelihood and the size of move in parameter space that we propose.  When we simulate parameter values at the start of the particle filter we simulate from $p(\theta|z)$. 
Thus if $Z$ is less informative, this will be a more diffuse distribution. Hence our sample of parameter values will be cover a larger region of the parameter space, and this will allow for potentially 
bigger proposed moves. However, the greater spread of parameter values we sample will mean a larger proportion of them will be in regions of low posterior probability, which will be wasteful. 
As a result the Monte Carlo variance of our estimate of the marginal-likelihood is likely to increase. Similar considerations will apply if we let $Z$ depend on $X_1$.

To gain some intuition about the trade-off in the choice of $Z$ we implemented particle MCMC for the linear-Gaussian model with $Z=(Z_x,Z_\theta)$ chosen as above. We simulated data as described in Section
\ref{S:LGM}, but with $\sigma_1=\sigma_\theta=1,000$. Our aim is to investigate how the performance of the new particle MCMC algorithm varies as we vary the variance of the noise in the definition of $Z_x$ and $Z_\theta$.
For this model we can calculate analytically the true posterior distribution for $X_1$ and $\theta$, and we chose the variance of the pseudo observations to be proportional to the marginal posterior variances. So for
a chosen $k_Z$ we set
\[
 \mbox{Var}(Z_x|x_1)=k_Z\mbox{Var}(X_1|y_{1:100}), ~~\mbox{and }~~ 
 \mbox{Var}(Z_\theta|\theta)=k_Z\mbox{Var}(\theta|y_{1:100}).
 \]
Figure \ref{Fig:LGSMM} (b) shows the resulting auto-correlation times for $X_1$ and $Z_X$ as we vary $k_Z$. Choosing $k_Z\approx 0$ gives auto-correlation times similar to running particle MCMC with 
$\mathcal{Z}=\{X_1,\theta\}$. As $k_Z$ is increased the efficiency of the particle MCMC algorithm initially increases, due to the better mixing of the underlying MCMC algorithm as we run particle MCMC 
conditioning on less information. However for very large $k_Z$ values the efficiency of particle MCMC becomes poor. In this case it starts behaving like particle MCMC with $\mathcal{Z}=\emptyset$, for
which the large Monte Carlo error in estimating the likelihood leads to poorer mixing. The best values of $k_Z$ correspond to adding noise to the pseudo-observations which is similar in size to the
marginal posterior variances of $X_1$ and $\theta$, and we notice good performance for a relatively large range of $k_Z$ values.

The improvement in mixing as we initially increase $k_Z$ is due to two aspects. The MCMC moves are updating $(Z_x,Z_\theta)$, and as we increase the noise for these pseudo-observations we reduce the posterior
dependence between them. Thus we observed a reduction in the auto-correlation time for $Z_x$. However, over and above this, we have that $\theta$ and $X_1$ are able to vary given values of $Z_x$ and $Z_\theta$. So for
larger noise in the pseudo-observations we see substantially smaller auto-correlation times for $X_1$ than for $Z_x$.

\subsection{MCMC within PMCMC}

One approach to improve the performance of a particle filter is to use MCMC moves within it \cite[see][]{Fearnhead:1998,Gilks/Berzuini:2001}. An example is to use a MCMC kernel to update particles prior to
propagating them to the next time-step. This involves a simple adaptation of Algorithm \ref{Alg:1}. Assume that $K_{t-1}(\cdot|\cdot)$ is a Markov kernel that has $p(\mathcal{X}_{t-1}|y_{1:t-1},\mathcal{Z})$ 
as its stationary distribution. Then 
we change step of 9 of Algorithm \ref{Alg:1} to:
\begin{itemize}
\item[9:] Sample $\mathcal{X}_{t-1}^*$ from $K_{t-1}(\cdot|\mathcal{X}_{t-1}^{(j)})$, and $\mathcal{X}_t^{(i)}$ from $p(\mathcal{X}_t|\mathcal{X}_{t-1}^*,\mathcal{Z})$.
\end{itemize}

The use of such an MCMC can be particularly helpful for updating parameters, as they help to ensure some diversity in the set of parameter values stored by the particles is maintained. Where possible, a common choice of 
kernel is to update just the parameters of the particle by sampling from the full conditional $p(\theta|x_{1:t-1},y_{1:t-1})$. Often such updates can be implemented in a computationally efficient manner as
the full conditional distribution just depends on the state-path through fixed-dimensional sufficient statistics \cite[]{Storvik:2002,Fearnhead:2002}. For recent examples of the benefits of using such MCMC moves see, 
for example, \cite{Carvalho:2010}, \cite{Carvalho:2010b} and \cite{Gramacy:2011}.

For standard implementations of particle MCMC, where $\mathcal{Z}=\theta$, using MCMC to update the parameters within the particle filter is not possible. 
Whereas by introducing pseudo-observations for the parameters, $Z$, and then implementing particle MCMC with $\mathcal{Z}=Z$ we can use such MCMC moves 
within the particle filter, or conditional particle filter. This can be of particular benefit if we use information from all particles, rather than just a single one.
\cite{Andrieu/Doucet/Holenstein:2010} suggest an approach for doing this using Rao-Blackwellisation idea. We consider an alternative approach in Section \ref{S:SV}.

\section{Examples} \label{S:EX}


\subsection{Stochastic Volatility} \label{S:SV}

A simple stochastic volatility model assumes a univariate state-process, defined as
\[
X_1=\sigma_0 \epsilon^{(X)}_1; \mbox{and } X_t=\gamma X_{t-1}+\sigma_X \epsilon^{(X)}_t, \mbox{ for $t=2,\ldots,T$},
\]
where $\epsilon^{(X)}_t$ are independent standard normal random variables. For $t=1,\ldots,T$ we have observations 
\[
 Y_t=\sigma_Y \exp\{x_t\} \epsilon^{(Y)}_t,
\]
where $\epsilon^{(Y)}_t$ are independent standard normal random variables. Thus the state process governs the variance of the observations, with larger values of $x_t$ meaning larger variability in the observation
at time $t$.

We assume $\theta=(\gamma,\sigma_X,\sigma_Y)$ are unknown. We introduce independent priors, with $\gamma$ having a normal distribution with mean $\mu_\gamma$ and variance $\sigma^2_\gamma$, but truncated to $(-1,1)$; while $\beta_X=1/\sigma_X^2$ and
$\beta_Y=1/\sigma_Y^2$ have gamma prior distributions with shape parameters $a_X$ and $a_Y$ respectively, and scale parameter $b_X$ and $b_Y$ respectively. We assume that $\sigma_0=1$.

We introduce a four-dimensional pseudo-observation $Z=(Z_X,Z_\gamma,Z_{\beta_X},Z_{\beta_Y})$ where conditional on $(X_1,\gamma,\beta_X,\beta_Y)$ 
\[
Z_X\sim \mbox{N}(X_1,\tau^2_X), ~~~ Z_\gamma \sim \mbox{N}(\gamma,\tau^2_\gamma), 
\]
\[
Z_{\beta_X}\sim \mbox{gamma}(n_X,\beta_X), \mbox{and } Z_{\beta_Y} \sim \mbox{gamma}(n_Y,\beta_Y).
\]
This choice for the pseudo-observations ensures that we can calculate the required marginal and conditional distributions, see Appendix \ref{App:A} for details.
To finalise the specification of these models we need to choose the values for $\tau_X$, $\tau_\gamma$, $n_X$ and $n_Y$ which determine how informative the pseudo-observations are.

{\bf Particle Gibbs}

We first compare different implementations of the Particle Gibbs algorithm. Our focus here is to show that using pseudo-observations can improve mixing in scenarios where the underlying
Gibbs sampler, even if it could be implemented, would mix poorly. For the stochastic volatility model this corresponds to situations where there is strong dependence in the state-process.

We simulated data with $T=1,000$ observations, $\gamma=0.99$, $\sigma_Y=1$ and $\sigma_X=1/(1-0.99^2)$, so that the stationary variance of the state process is 1. We present results for priors
with $\mu_\gamma=0.5$ and $\sigma^2_\gamma=0.5$; $a_X=1$ and $b_X=1/1000$; and $a_Y=0.1$ and $b_Y=0.1$. This corresponds to the true values of $\gamma$ and $\beta_X$ being in the tails of the prior, and a relatively
uninformative prior for $\beta_Y$.

We implemented both the standard version of Particle Gibbs, with $\mathcal{Z}=\theta$, and Particle Gibbs with conditioning on the pseudo-observations, 
$\mathcal{Z}=Z$ defined above. We chose the tuning parameters of the pseudo observations so that the variance of the parameters given $Z$ was slightly smaller than the posterior variance we observed from 
a pilot run. For further comparison we show results for Particle Gibbs with no conditioning, $\mathcal{Z}=\emptyset$, again implemented with $N=500$.

We ran the standard version with $N=250$ particles, and the other two versions with $N=500$. This was based on choosing $N$ so that the estimate of the
log-likelihood had a variance of around 1 \cite[]{Pitt:2012}. To compensate for the doubling of the computational cost of
the conditional SMC sampler with the latter two versions, we ran the standard version of the Particle Gibbs for twice as many iterations. To ease comparison of results we then thinned the output by keeping
the values of the chain on even iterations only. 

\begin{figure}
\begin{center}
 \includegraphics[scale=0.5]{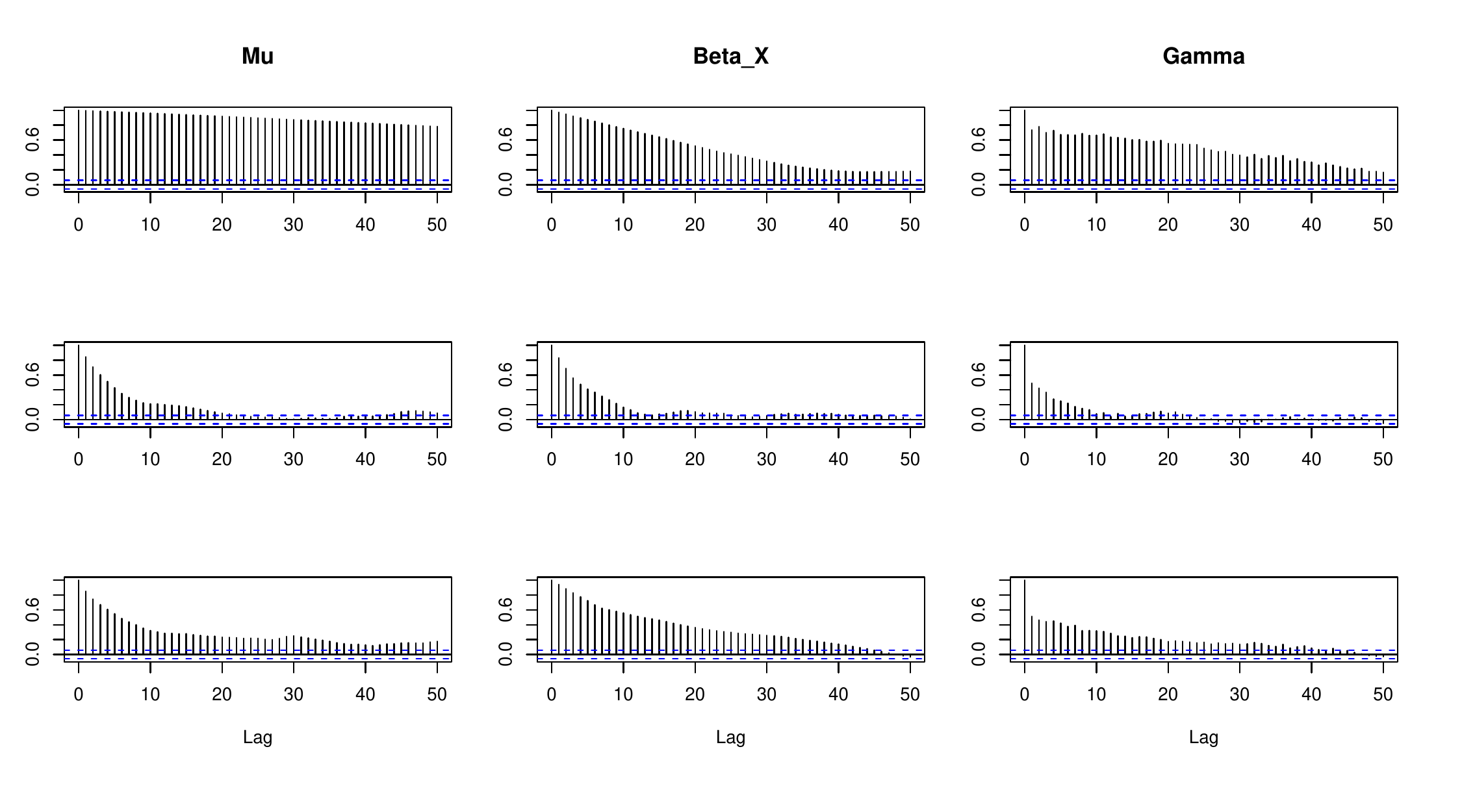}
 \caption{\label{Fig:SV1} ACF plots for three runs of the Particle Gibbs conditioning on: $\mathcal{Z}=\theta$ (top row); $\mathcal{Z}=Z$ (middle row); and $\mathcal{Z}=\emptyset$ (bottom row). Each column corresponds
 to ACF for a different parameter: $\mu=\log(\beta_Y)$ (left column); $\beta_X$ (middle column); and $\gamma$ (right column).}
\end{center}
\end{figure}

Results are shown in Figure \ref{Fig:SV1}. The standard implementation performs badly here. This is because of strong dependencies between the parameters and the state-process that occurs for this model which means
that the underlying Gibbs sampler mixes slowly. By conditioning on less information when running the conditional SMC sampler we reduce this dependence between the $\mathcal{X}_T$ and $\mathcal{Z}$ which improves mixing. 
However, choosing $\mathcal{Z}=\emptyset$ results in a substantial decrease in efficiency of the conditional SMC sampler. This is particularly pronounced due to the relatively uninformative priors we chose, and the
fact that one of the parameter values was in the tail of the prior. If much more informative priors were chosen, using $\mathcal{Z}=\emptyset$ would give similar results to the use of pseudo-observations.
Also this effect could be reduced slightly by increasing $N$ further for this implementation of 
Particle Gibbs, but doing so will still lead to a less efficient sampler than using pseudo-observations. 

Finally, we note that there are alternative approaches to reduce correlation for a Gibbs sampler, such as reparameterisation approaches \cite[]{Pitt/Shephard:1999}. Getting such ideas to work often requires
implementations that are specific to a given application. By comparison, the use of pseudo-observations gives
 a general way of reducing the correlation between $\mathcal{X}_T$ and $\mathcal{Z}$ that adversely affects the mixing of the underlying Gibbs sampler.

{\bf PMMH with MCMC}

We now compare PMMH on the stochastic volatility model. Our focus is purely on how using MCMC within the particle filter can help improve mixing over a standard PMMH algorithm. We simulated data with parameter values
as above. To help reduce the computational cost involved in analysing this data, and hence implementing the simulation study, using PMMH we use more informative priors (which meant we could use fewer particles when running the particle filters),
with $\mu_\gamma=0.9$ and $\sigma^2_\gamma=0.1$; $a_X=1$ and $b_X=1/100$; and $a_Y=1$ and $b_Y=1$.

We compared two implementations of PMMH, one with $\mathcal{Z}=\theta$ and one with $\mathcal{Z}=Z$. For the latter we were able to use MCMC within the particle filter to update the parameter values, using
standard Particle Learning algorithms \cite[]{Carvalho:2010}. Using the criteria of \cite{Pitt:2012}, we chose $N=150$ and $N=450$ particles respectively for these implementations. We
used random walk proposals with the variances informed by a pilot run \cite[]{Roberts/Rosenthal:2001}. Again we compensate for the slow running of the PMMH with pseudo-observations by running the other PMMH algorithm for
three times as long, and thinning: keeping only every third value. 

The main improvement in effiency we observed with the second PMMH algorithm was through using the diversity in parameter values we obtain when using the Particle Learning Algorithm. Our approach for implementing this was to
output a set of equally weighted particle values from the particle learning algorithm. We then make a decision as to whether to accept this set of particles, with the normal acceptance probability. Finally, we add
an extra step to each iteration where we resample the state of the PMMH algorithm from the last stored set of particles. 
Full details are given in Algorithm \ref{Alg:4}. 

\begin{algorithm}
   \caption{Particle marginal Metropolis-Hastings Algorithm with Particle Learning}
   \label{Alg:4}
 {\bf Input:} \\
An initial value $\mathcal{Z}^{(0)}$. \\
A proposal distribution $q(\cdot|\cdot)$.\\
The number of particles, $N$, and the number of MCMC iterations, $M$.
  \begin{algorithmic}[1]
\STATE Run a Particle Learning Algorithm  with $N$ particles, conditioning on $\mathcal{Z}^{(0)}$, to obtain a set of equally weighted particle $\{\mathcal{X}_T^{(0,j)}\}_{j=1}^N$ and $\hat{p}(y_{1:T}|\mathcal{Z}^{(0)})$.
\STATE Obtain $\mathcal{X}_T^0$ by sampling uniformaly at random from $\{\mathcal{X}_T^{(0,j)}\}_{j=1}^N$
     \FOR{$i=1,\ldots,M$}  
      \STATE Sample $\mathcal{Z}'$ from $q(\mathcal{Z}|\mathcal{Z}^{(i-1)})$.
     \STATE Run a Particle Learning Algorithm with $N$ particles, conditioning on $\mathcal{Z}'$, to obtain a set of equally weighted particle $\{\mathcal{X}_T^{(*,j)}\}_{j=1}^N$ and $\hat{p}(y_{1:T}|\mathcal{Z}')$. 
    \STATE With probability
\[
\min\left\{ 1, \frac{q(\mathcal{Z}^{(i-1)}|\mathcal{Z}')\hat{p}(y_{1:T}|\mathcal{Z}')p(\mathcal{Z}')}{q(\mathcal{Z}'|\mathcal{Z}^{(i-1)})\hat{p}(y_{1:T}|\mathcal{Z}^{(i-1)})p(\mathcal{Z}^{(i-1)})} \right\}
\]
set $\{\mathcal{X}_T^{(i,j)}\}_{j=1}^N=\{\mathcal{X}_T^{(*,j)}\}_{j=1}^N$ and $\hat{p}(y_{1:T}|\mathcal{Z}^{(i)})=\hat{p}(y_{1:T}|\mathcal{Z}')$; otherwise set $\{\mathcal{X}_T^{(i,j)}\}_{j=1}^N=\{\mathcal{X}_T^{(i-1,j)}\}_{j=1}^N$
 and $\hat{p}(y_{1:T}|\mathcal{Z}^{(i)})=\hat{p}(y_{1:T}|\mathcal{Z}^{(i-1)})$
 \STATE Obtain $\mathcal{X}_T^i$ by sampling uniformaly at random from $\{\mathcal{X}_T^{(i,j)}\}_{j=1}^N$
      \ENDFOR
\end{algorithmic}
{\bf Output:} A sample of extended state vectors: $\{\mathcal{X}_T^{(i)}\}_{i=1}^M$.
\end{algorithm}

\begin{figure}
\begin{center}
 \includegraphics[scale=0.5]{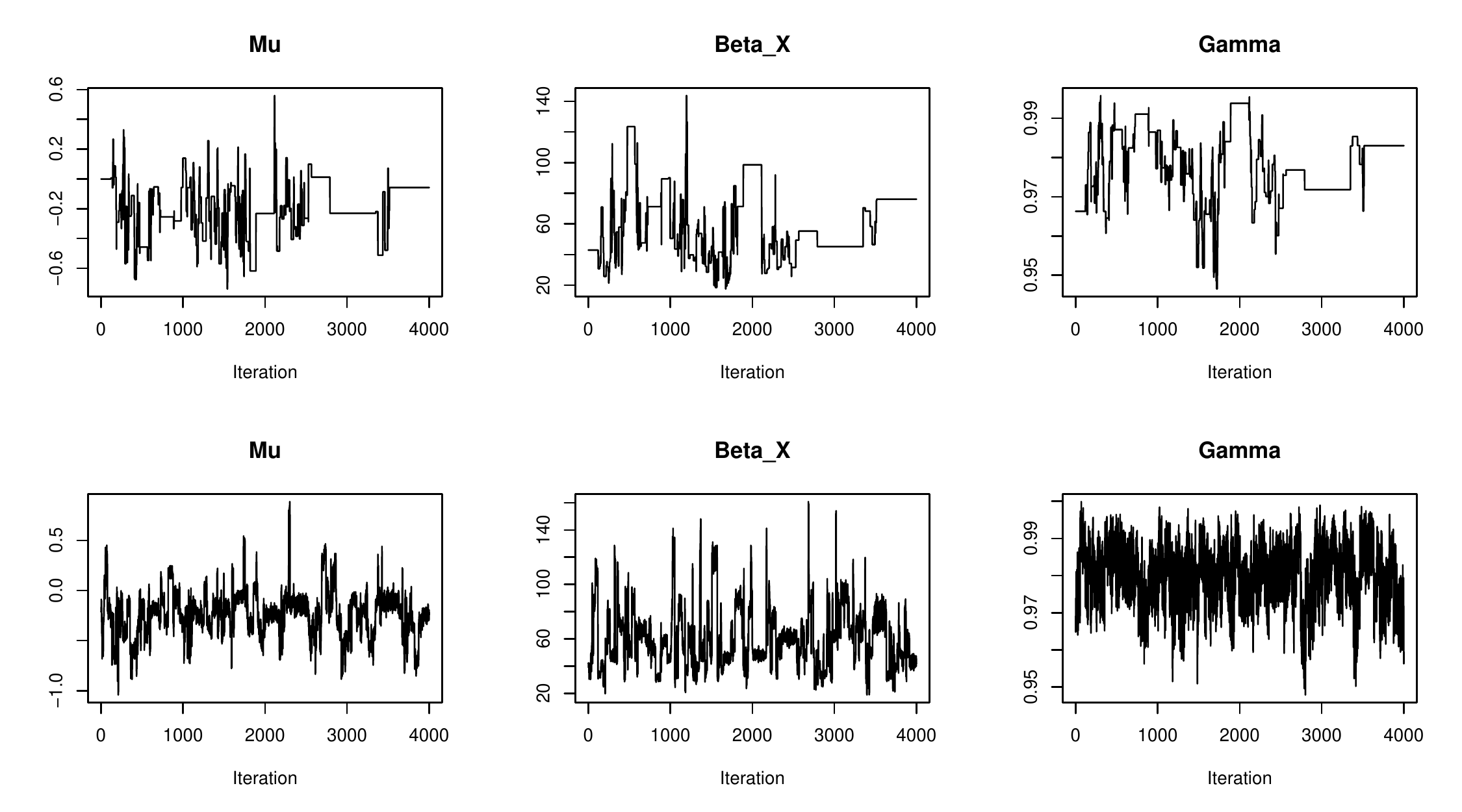}
 \caption{\label{Fig:SV2} Trace plots for two runs of PMMH: $\mathcal{Z}=\theta$ (top row); and $\mathcal{Z}=Z$ (bottom row). Each column corresponds
 to a different parameter: $\mu=\log(\beta_Y)$ (left column); $\beta_X$ (middle column); and $\gamma$ (right column).}
\end{center}
\end{figure}

Trace-plots from part of the PMMH run are shown in Figure \ref{Fig:SV2}. These highlight the main improvement that using particle learning within PMMH gives. Both runs of PMMH can have long periods were they reject the output
of the particle filter. However, by utilising the diversity in the parameter values of the particles that are output when particle learning is used, the PMMH algorithm is still able to mix over different parameter
values in that case. Calculations of effective sample sizes show that this leads to a roughly three-fold increase in effective sample sizes (for a given CPU cost) for estimating $\beta_X$ and $\gamma$.

Remember we cannot use particle learning to update parameters for the standard implementation of PMMH, where $\mathcal{Z}=\theta$, as in that case the particle filter is implemented conditional
on a fixed set of parameter values. 

\subsection{Dirichlet Process Mixture Models} \label{S:DPMM}

We now consider inference for a mixture model used to infer population structure from population genetic data. Assume we have
data from a set of diploid individuals, and this data consists of the genotype of each individual at a set of unlinked loci. Thus each
locus will have a set of possible alleles (different genetic types), and the data for an individual at that locus will be which alleles
are present on each of two copies of that individual's genome. We further assume that the individuals each come from one of an unknown
number of populations. The frequency of each allele at each locus will vary across these populations. We wish to infer how many populations
there are, and which individuals come from the same population.

This is an important problem in population genetics. We will consider a model based on that of \cite{Pritchard/Stephens/Donnelly:2000}. Though see \cite{Pritchard/Stephens/Donnelly:2000}, \cite{Nicholson:2002} and 
\cite{Falush/Stephens/Pritchard:2003} for extensions of this
model; \cite{Falush:2003}, \cite{Pritchard/Stephens/Rosenberg/Donnelly:2000} and \cite{Rosenberg:2002} for example applications; and \cite{Price:2006} and \cite{Patterson:2006} for alternative approaches to this problem.

Assume we have $L$ loci. At locus $l$ we have $K_l$ alleles. The allele frequencies of these alleles in population $j$ are given by $\mathbf{p}^{(j,l)}=(p_1^{(j,l)},\ldots,p_{K_l}^{(j,l)})$. The genotype of individual $i$ at 
locus $l$ is $\mathbf{y}_{i,l}=(y_{i,l}^{(1)},y_{i,l}^{(2)})$. Let $x_i$ be a unobserved latent variable which defines the population that individual $i$ is from. Then the conditional likelihood of 
$\mathbf{y}_i=(\mathbf{y}_{i,1},\ldots,\mathbf{y}_{i,L})$ given $x_i=j$ is
\[
 p(\mathbf{y}_i|x_i=j)=\prod_{l=1}^L p_{y_{i,l}^{(1)}}^{(j,l)}p_{y_{i,l}^{(2)}}^{(j,l)}.
\]
This model assumes the loci are unlinked and there is no admixture, hence conditional on $x_i$ the data at each locus are independent.

We assume conjugate Dirichlet priors for the allele frequencies in each population. These priors are independent across both loci and population. For locus $l$ the parameter vector of the Dirichlet prior
is $(\lambda/K_{l},\ldots,\lambda/K_{l})$.

We use a mixture Dirichlet process (MDP) model \cite[]{Ferguson:1973} for the prior distribution of latent variables $x_i$.
We will use the following recursive representation of the MDP model \cite[]{Blackwell/MacQueen:1973}. 
Let ${x}_{1:i}=(x_1,\ldots,x_i)$ be the population of origin of the first $i$ individuals, and define $m({x}_{1:i})$ to be the number of populations present in ${x}_{1:i}$. 
We number these populations $1,\ldots,m({x}_{1:i})$, and let $n_j({x}_{1:i})$ be the number of these individuals assigned to population $j$. Then
\begin{equation} \label{eq:DPMM}
 p(x_{i+1}=j|{x}_{1:i})=\left\{\begin{array}{cl} n_j({x}_{1:i})/(i+\alpha) & \mbox{if $j\leq m(x_{1:i})$}, \\
\alpha/(i+\alpha) & \mbox{if $j= m(x_{1:i})+1$}. \end{array}\right.
\end{equation}
This model does not pre-specify the number of populations present in the data. Note that the actual labelling of populations under the MDP model is arbitrary, and the information in $x_{1:n}$ is essentially which
subset of individuals belong to each of the populations. In our implementation the actual labels are defined by the order of the individuals in the data set. With population 1 being the population that the first 
individual belongs to, population 2 is the population that the first individual not in population 1 belongs to, and so on.

Inference for this model was considered in \cite{Fearnhead:2008} for the case where $\lambda$ and $\alpha$ were known. Here we introduce hyperpriors for both these parameters, and perform inference using particle MCMC. We use independent
gamma priors, with $\alpha\sim\mbox{Gamma}(5,10)$ and $\lambda\sim\mbox{Gamma}(4,1)$. If we condition on values for $\lambda$ and $\alpha$, then \cite{Fearnhead:2008} presents an efficient particle algorithm for this problem. 
This particle filter is based on ideas in \cite{Fearnhead/Clifford:2003} and \cite{Fearnhead:2004SC1}. 

\begin{figure}
\begin{center}
 \includegraphics[scale=0.9]{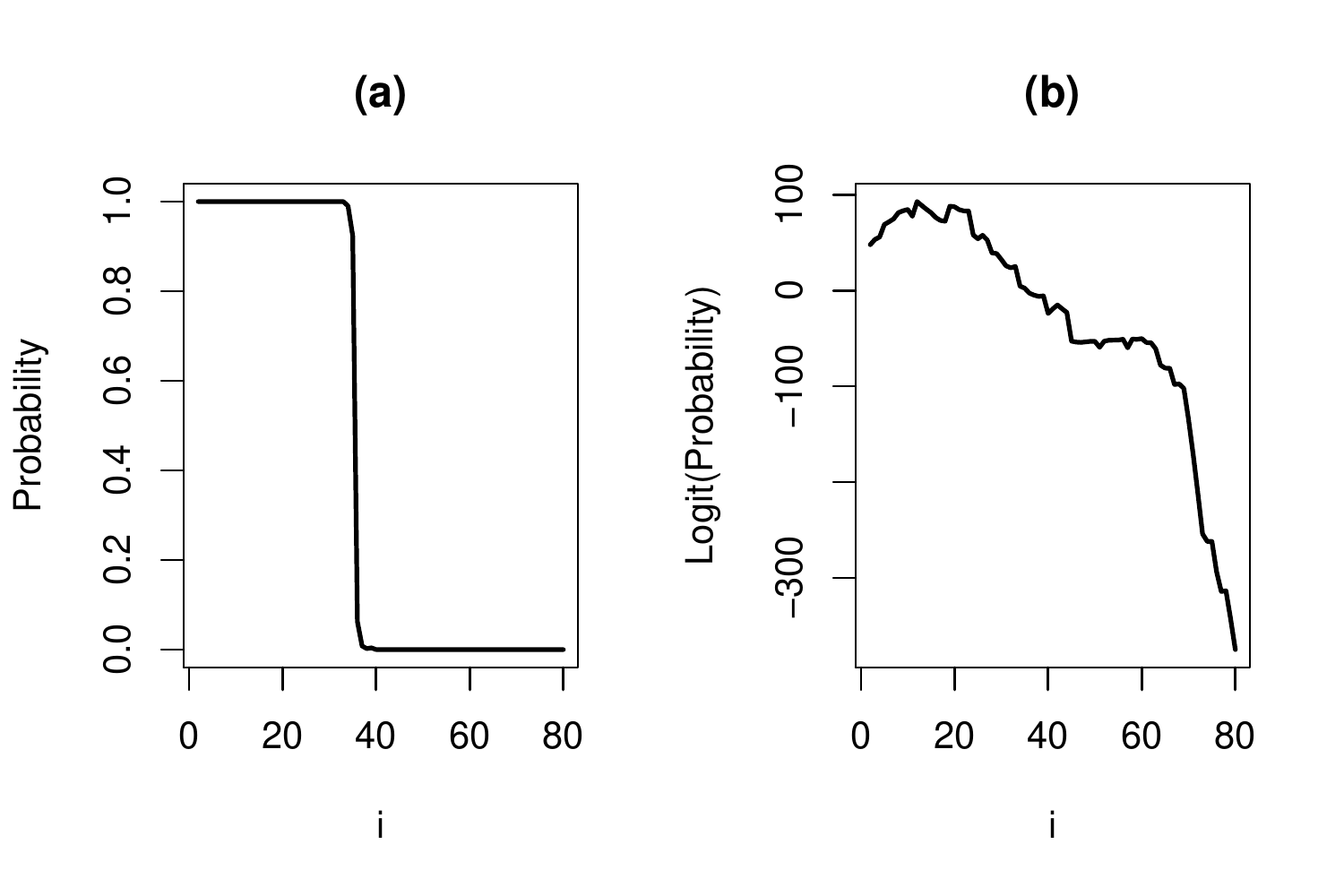}
 \caption{\label{Fig:1} Plot of (a) $\Pr(X_1=X_2|y_{1:i})$  and (b) $\mbox{logit}[\Pr(X_1=X_2|y_{1:i})]$ for different sizes of data set $i$.}
\end{center}
\end{figure}

The particle filter of \cite{Fearnhead:2008} can struggle in applications where $L$ is large, due to problems with initialisation. To show this we considered inference for $n=80$ individuals at $L=100$ loci, 
using a subset of data taken from \cite{Rosenberg:2002}. Figure \ref{Fig:1} plots estimates of $\Pr(X_1=X_2|y_{1:i})$ for increasing values of $i$. This shows
how the posterior probability of the first two individuals being from the same population changes as we analyse data from more people. Initially this is close to 1, whereas once all data has been analysed the probability is 
essentially 0. This substantial change in probability causes problems in a particle filter, as all particles with $x_1\neq x_2$ are likely to lost during resampling in the early iterations of the algorithm. 

To overcome this problem of initialisation of the particle filter for this application we propose to introduce a pseudo observation, $Z_x$, that contains information about the populations of a random subset of the individuals.
The distribution of $Z_x$ given $x_{1:n}$ is obtained by (i) sampling the number of individuals in the subset, $v$ say; (ii) choosing $v$ individuals at random from the sample, $i_1,\ldots,i_v$; and (iii) 
letting $Z_x=\{(i_1,x_{i_1}),\ldots,(i_v,x_{i_v})\}$, the subset of individuals and their population labels.

As mentioned above, the actual values of the population labels is arbitrary, and $Z_x$ just contains information about which of the individuals $i_1,\ldots,i_v$ belong to the same population. In practice, at each
iteration we re-order the individuals in the sample so that individuals $i_1,\ldots,i_v$ become the first $v$ individuals, and the order of the remaining individuals is chosen uniformly at random. The labels for the new first
$v$ individuals are changed to be consistent with our recursive represenation of the MDP model above.

In implementing PMCMC we use $\mathcal{Z}=(\lambda,\alpha,Z_x)$. Our proposal distribution for $Z_x$ is just its true conditional distribution given $x_{1:n}$. 
We can easily adapt the particle filter of \cite{Fearnhead:2008} to condition on $\mathcal{Z}$, 
by fixing the labels of the first $v$ individuals in the sample to those specified by $Z_x$. We use a random walk proposal for updating $\log \lambda$ and an independence proposal for $\alpha$. Further details
are given in Appendix \ref{App:B}.

\begin{figure}
\begin{center}
 \includegraphics[scale=0.6]{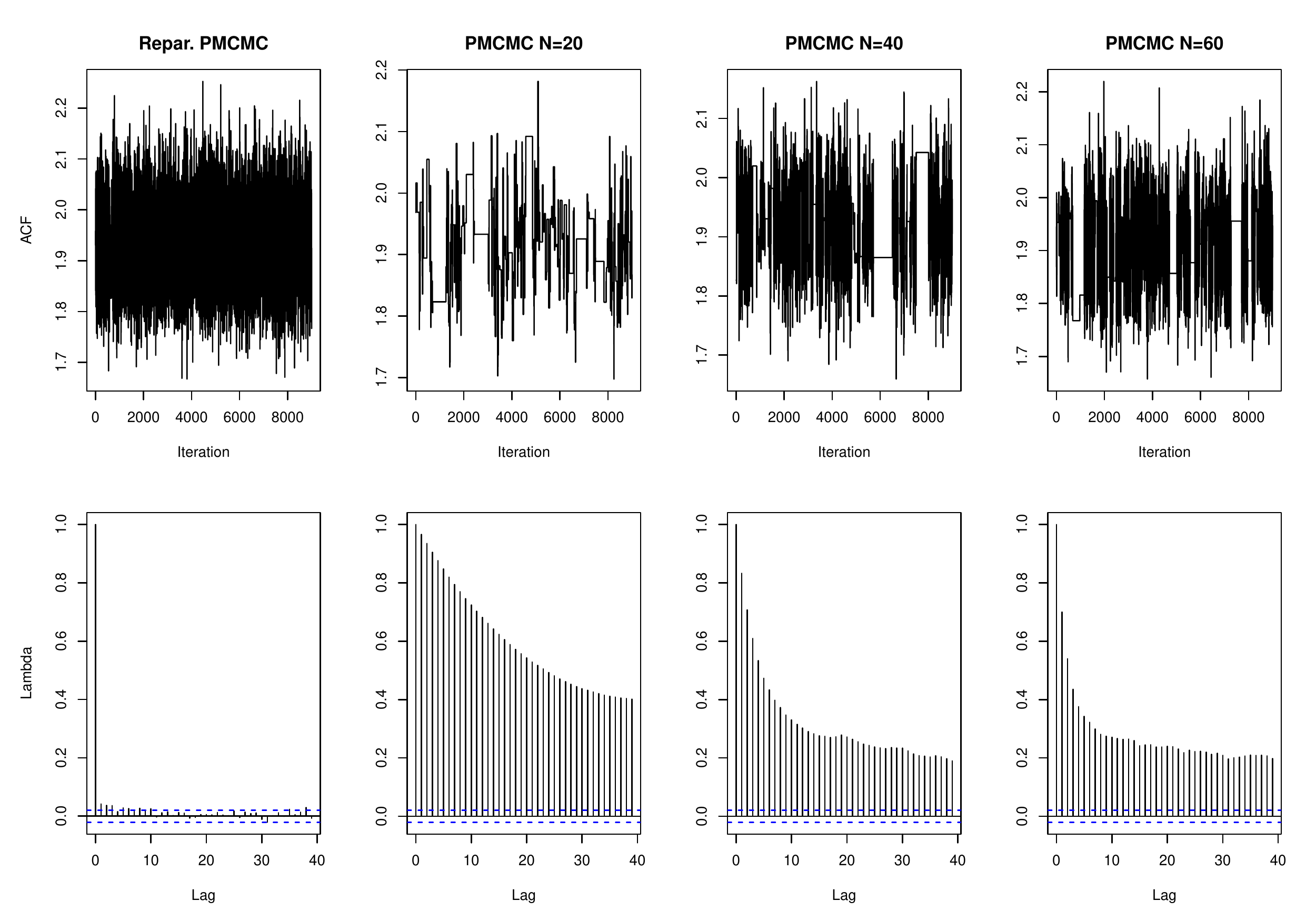}
 \caption{\label{Fig:2} Trace plots (top) and acf plots (bottom) for $\lambda$ for the reparameterised PMCMC with $Z=(\lambda,\alpha,Z_x)$ (left-hand column), and standard PMCMC with $Z=(\lambda,\alpha)$ (other columns). We ran the reparameterised PMCMC
 with $N=20$ particles, and the standard PMCMC with $N=20$, $N=40$ and $N=60$ particles. Results shown after removing the first $10^5$ iterations as burn-in and thinning the remaining output by keeping only every 100th value.}
\end{center}
\end{figure}

We compared the reparameterised PMCMC with this choice of $Z$ with a standard PMCMC algorithm where $Z=(\lambda,\alpha)$. Our aim is purely to investigate the relative efficiency of the two implementations of PMCMC on this challenging problem. We
ran each PMCMC algorithm for $10^6$ iterations, storing only every 100th value. We implemented the new PMCMC algorithm using 20 particles for the particle filter, and with $Z_x$ storing population 
information from an average of 5 individuals. We implemented the standard PMCMC algorithm with 20, 40 and 60 particles. Results, in terms of trace and acf plots for $\lambda$ are shown in Figure \ref{Fig:2}.

We see that the reparameterised PMCMC algorithm has substantially better mixing than the standard PMCMC algorithm, even when the latter used 3 times as many particles, and hence would have three times the CPU cost per iteration.
For all the standard PMCMC algorithms, the chain gets stuck for substantial periods of time. This is due to a large variance of the estimate of the likelihood. By running the particle filter conditional on $Z_x$ we 
obtain a substantial reduction in the variance of our estimates of the likelihood, and hence avoid this problem.  

Estimated auto-correlation times are 1.3 for the reparameterised PMCMC algorithm with 20 particles, 
and 105, 56 and 36 for the standard PMCMC with 20, 40 and 60 particles respectively. After taking account that the CPU cost of an iteration of PMCMC is proportional to the number of particles, 
this suggests the re-parameterised PMCMC is about 80 times more efficient than each of the standard PMCMC algorithms.

\section{Discussion}

We have  introduced a way to generalise particle MCMC through data augmentation. The idea is to introduce new latent variables into the model, and then to implement particle MCMC where the MCMC moves update the
latent variables, and the particle filter updates the rest of the variables in the model. By careful choice of the latent variables, we have shown this can lead to substantial gains in efficiency in situations
where the standard particle MCMC algorithm performs poorly. For the Stochastic Volatility example of Section \ref{S:SV} we saw that it can help break down dependencies that make the particle Gibbs algorithm mix slowly,
and can enable particle learning ideas to be used within the particle filter component of particle MCMC. It can also help for models where the particle filter struggles with initialisation, that is where
at early time-steps the filter is likely to
sample particles in areas that are inconsistent with the full data, as we saw in Section \ref{S:DPMM}.

The ideas in this paper bear some similarity with the marginal augmentation approaches for improving the Gibbs sampler \cite[e.g.][]{vanDyk:2001}. In both cases, adding a latent variable to the model, and implementing the MCMC
algorithm for this expanded model, can improve mixing. Our way of introducing the latent variables, and the way they are used are completely different though. However, both approaches can improve mixing for the same
reason. Introducing the latent variables reduces the correlation between variables updated at different stages of the Gibbs, or particle Gibbs, sampler.

The new data augmentation ideas add great flexibility to the particle MCMC algorithm. One key open question is how to choose the best latent variables to introduce, or, equivalently, how to tune the variance of the 
pseudo-observations. Our experience to date suggests that you want to choose this so that the conditional distribution of the parameters, or of the initial state, given the pseudo-observations has a similiar variance to 
that of the posterior distribution. 

{\bf Acknowledgements:} The first author was supported by the Engineering and Physical Sciences Research Council grant EP/K014463/1.

\appendix
\section{Calculations for the Stochastic Volatility Model} \label{App:A}

First consider $Z_{\beta_X}$. Standard calculations give
\begin{eqnarray*}
 p(\beta_X|z_{\beta_X}) &\propto& p(\beta_X)p(z_{\beta_X}|\beta_X) \\
&\propto& \beta_x^{a_x-1}\exp\{-b_x\beta_x\}\left( \beta_X^{n_x}\exp\{-\beta_x z_{\beta_X}\}\right)
\end{eqnarray*}
This gives that the conditional distribution of $\beta_X$ given $z_{\beta_X}$ is gamma with parameters $n_x+a_x$ and $b_x+z_{\beta_X}$. Furthermore the
marginal distribution for $Z_{\beta_X}$ is
\begin{eqnarray*}
 p(z_{\beta_X}) &=& \int p(\beta_X)p(z_{\beta_X}|\beta_X)\mbox{d}\beta_X \\
 &=& \frac{b_x^{a_x}z_{\beta_x}^{n_x-1}}{\Gamma(a_x)\Gamma(n_x)}\int  \beta_x^{a_x+n_x-1}\exp\{-(b_x+z_{\beta_X}\beta_x\}     \mbox{d}\beta_X \\
 &=& \left(\frac{\Gamma(a_x+n_x)b_x^{a_x}}{\Gamma(a_x)\Gamma(n_x)} \right)\left(\frac{z_{\beta_x}^{n_x-1}}{(z_{\beta_x}+b_x)^{n_x+a_x}} \right)
\end{eqnarray*}
The calculations for $Z_{\beta_Y}$ are identical. 

Calculations for $Z_X$ and $Z_\gamma$ are as for the linear Gaussian model (see Section \ref{S:EAS}).

\section{Calculations for the Dirichlet Process Mixture Model} \label{App:B}

The conditional distribution of $Z_x$ given $x_{1:n}$ can be split into (i) the marginal distribution for $v$, $p(v)$;
(ii) the conditional distribution of the sampled individuals, $i_1,\ldots,i_v$, given $v$. Given $i_1,\ldots,i_v$, the
clustering of these individuals is deterministic, being defined by the clustering $(x_{i_1},\ldots,x_{i_v})$.

The marginal distribution of $Z_x$ thus can be written as
\[
 p(Z_x)=p(v)p(i_1,\ldots,i_v|v)p(x_{i_1},\ldots,x_{i_v}).
\]
Where we that, due to uniform sampling of the individuals, \[p(i_1,\ldots,i_v|v)=\left(\begin{array}{c} n \\ v \end{array} \right).\]
Finally $p(x_{i_1},\ldots,x_{i_v})$ is given by the Dirichlet process prior. If we relabel the populations so that $x_{i_1}=1$,
population 2 is the population of the first individual in $i_1,\ldots,i_v$ that is not in population 1, and so on; then for $v>1$, 
\[
 p(x_{i_1},\ldots,x_{i_v})=\prod_{j=2}^v p(x_{i_j}|x_{i_1},\ldots,x_{i_{j-1}}),
\]
with $p(x_{i_j}|x_{i_1},\ldots,x_{i_{j-1}})$ defined by (\ref{eq:DPMM}).

Within the PMMH we use a proposal for $Z_x$ given $X_{1:n}$ that is its full conditional
\[
 q(Z_x|x_{1:n})=p(Z_x|x_{1:n})=p(v)p(i_1,\ldots,i_v|v).
\]
In practice we take the distribution of $v$ to be a Poisson distribution with mean 5, truncated to take values less than $n$. (Similar results
were observed as we varied both the distribution and the mean value.)

\bibliographystyle{royal}

\bibliography{/home/fearnhea/bib/coal,/home/fearnhea/bib/thesis}

\begin{thebibliography}{39}
\expandafter\ifx\csname natexlab\endcsname\relax\def\natexlab#1{#1}\fi

\bibitem[Andrieu and Roberts(2009)]{Andrieu/Roberts:2009}
Andrieu, C. and Roberts, G.~O. (2009). The pseudo-marginal approach for
  efficient computations. {\em Annals of Statistics\/} {\bf 37}, 697--725.

\bibitem[Andrieu {\em et~al.\/}(2010)Andrieu, Doucet and
  Holenstein]{Andrieu/Doucet/Holenstein:2010}
Andrieu, C., Doucet, A. and Holenstein, R. (2010). {Particle Markov chain Monte
  Carlo (with Discussion)}. {\em Journal of the Royal Statistical Society,
  Series B\/} {\bf 62}, 269--342.

\bibitem[Blackwell and Mac{Q}ueen(1973)]{Blackwell/MacQueen:1973}
Blackwell, D. and Mac{Q}ueen, J.~B. (1973). Ferguson distributions via {P}olya
  urn schemes. {\em Annals of Statistics\/} {\bf 1}, 353--355.

\bibitem[Carpenter {\em et~al.\/}(1999)Carpenter, Clifford and
  Fearnhead]{Carpenter/Clifford/Fearnhead:1999}
Carpenter, J., Clifford, P. and Fearnhead, P. (1999). An improved particle
  filter for non-linear problems. {\em IEE proceedings-Radar, Sonar and
  Navigation\/} {\bf 146}, 2--7.

\bibitem[Carvalho {\em et~al.\/}(2010{\natexlab{a}})Carvalho, Johannes, Lopes
  and Polson]{Carvalho:2010}
Carvalho, C.~M., Johannes, M.~S., Lopes, H.~F. and Polson, N.~G.
  (2010{\natexlab{a}}). Particle learning and smoothing. {\em Statistical
  Science\/} , 88--106.

\bibitem[Carvalho {\em et~al.\/}(2010{\natexlab{b}})Carvalho, Lopes, Polson and
  Taddy]{Carvalho:2010b}
Carvalho, C.~M., Lopes, H.~F., Polson, N.~G. and Taddy, M.~A.
  (2010{\natexlab{b}}). Particle learning for general mixtures. {\em Bayesian
  Analysis\/} {\bf 5}, 709--740.

\bibitem[{Del Moral}(2004)]{DelMoral:2004}
{Del Moral}, P. (2004). {\em Feynman-Kac Formulae: Genealogical and Interacting
  Particle Systems With Applications\/}. Springer, New York.

\bibitem[Doucet {\em et~al.\/}(2000)Doucet, Godsill and
  Andrieu]{Doucet/Godsill/Andrieu:2000}
Doucet, A., Godsill, S.~J. and Andrieu, C. (2000). On sequential {Monte Carlo}
  sampling methods for {B}ayesian filtering. {\em Statistics and Computing\/}
  {\bf 10}, 197--208.

\bibitem[{Doucet} {\em et~al.\/}(2012){Doucet}, {Pitt} and {Kohn}]{Doucet:2012}
{Doucet}, A., {Pitt}, M. and {Kohn}, R. (2012). {Efficient implementation of
  Markov chain Monte Carlo when using an unbiased likelihood estimator}. {\em
  ArXiv e-prints\/} .

\bibitem[van Dyk and Meng(2001)]{vanDyk:2001}
van Dyk, D.~A. and Meng, X.-L. (2001). The art of data augmentation. {\em
  Journal of Computational and Graphical Statistics\/} {\bf 10}(1), 1--50.

\bibitem[Falush {\em et~al.\/}(2003{\natexlab{a}})Falush, Stephens and
  Pritchard]{Falush/Stephens/Pritchard:2003}
Falush, D., Stephens, M. and Pritchard, J.~K. (2003{\natexlab{a}}). Inference
  of population structure using multilocus genotype data: Linked loci and
  correlated allele frequencies. {\em Genetics\/} {\bf 164}, 1567--1587.

\bibitem[Falush {\em et~al.\/}(2003{\natexlab{b}})Falush, Wirth, Linz,
  Pritchard, Stephens, Kidd, Blaser, Graham, Vacher, Perez-Perez, Yamaoka,
  Megraud, Otto, Reichard, Katzowitsch, Wang, Achtman and
  Suerbaum]{Falush:2003}
Falush, D., Wirth, T., Linz, B., Pritchard, J.~K., Stephens, M., Kidd, M.,
  Blaser, M.~J., Graham, D.~Y., Vacher, S., Perez-Perez, G.~I., Yamaoka, Y.,
  Megraud, F., Otto, K., Reichard, U., Katzowitsch, E., Wang, X.~Y., Achtman,
  M. and Suerbaum, S. (2003{\natexlab{b}}). Traces of human migrations in
  {Helicobacter pylori} populations. {\em Science\/} {\bf 299}, 1582--1585.

\bibitem[Fearnhead(1998)]{Fearnhead:1998}
Fearnhead, P. (1998). {\em Sequential Monte Carlo methods in filter theory\/}.
  Ph.D. thesis, Oxford Unversity, available from
  {\texttt{http://www.maths.lancs.ac.uk/$\sim$fearnhea/}}.

\bibitem[Fearnhead(2002)]{Fearnhead:2002}
Fearnhead, P. (2002). {MCMC}, sufficient statistics and particle filters. {\em
  Journal of Computational and Graphical Statistics\/} {\bf 11}, 848--862.

\bibitem[Fearnhead(2004)]{Fearnhead:2004SC1}
Fearnhead, P. (2004). Particle filters for mixture models with an unknown
  number of components. {\em Statistics and Computing\/} {\bf 14}, 11--21.

\bibitem[Fearnhead(2008)]{Fearnhead:2008}
Fearnhead, P. (2008). Computational methods for complex stochastic systems: A
  review of some alternatives to {MCMC}. {\em Statistics and Computing\/} {\bf
  18}, 151--171.

\bibitem[Fearnhead(2011)]{Fearnhead:2011}
Fearnhead, P. (2011). Mcmc for state-space models. In: {\em Handbook of {Markov
  chain Monte Carlo}\/} (eds. S.~Brooks, A.~Gelman, G.~L. Jones and X.~Meng),
  Chapman \& Hall/CRC.

\bibitem[Fearnhead and Clifford(2003)]{Fearnhead/Clifford:2003}
Fearnhead, P. and Clifford, P. (2003). Online inference for hidden {M}arkov
  models. {\em Journal of the Royal Statistical Society, Series B\/} {\bf 65},
  887--899.

\bibitem[Ferguson(1973)]{Ferguson:1973}
Ferguson, T.~S. (1973). A {B}ayesian analysis of some nonparametric problems.
  {\em Annals of Statistics\/} {\bf 1}, 209--230.

\bibitem[Gilks and Berzuini(2001)]{Gilks/Berzuini:2001}
Gilks, W.~R. and Berzuini, C. (2001). Following a moving target - {Monte Carlo}
  inference for dynamic {Bayesian} models. {\em Journal of the Royal
  Statistical Society, Series B\/} {\bf 63}, 127--146.

\bibitem[Golightly and Wilkinson(2011)]{Golightly/Wilkinson:2011}
Golightly, A. and Wilkinson, D.~J. (2011). Bayesian parameter inference for
  stochastic biochemical network models using particle markov chain monte
  carlo. {\em Interface Focus\/} .

\bibitem[Gramacy and Polson(2011)]{Gramacy:2011}
Gramacy, R.~B. and Polson, N.~G. (2011). Particle learning of gaussian process
  models for sequential design and optimization. {\em Journal of Computational
  and Graphical Statistics\/} {\bf 20}, 102--118.

\bibitem[Liu and Chen(1998)]{Liu/Chen:1998}
Liu, J.~S. and Chen, R. (1998). {Sequential Monte Carlo methods for dynamic
  systems}. {\em Journal of the American Statistical Association.\/} {\bf 93},
  1032--1044.

\bibitem[{Murray} {\em et~al.\/}(2012){Murray}, {Jones} and
  {Parslow}]{Murray:2012}
{Murray}, L.~M., {Jones}, E.~M. and {Parslow}, J. (2012). {On Disturbance
  State-Space Models and the Particle Marginal Metropolis-Hastings Sampler}.
  {\em ArXiv e-prints\/} .

\bibitem[Nicholson {\em et~al.\/}(2002)Nicholson, Smith, J\'{o}nsson,
  G\'{u}stafsson, Stef\'{a}nsson and Donnelly]{Nicholson:2002}
Nicholson, G., Smith, A.~V., J\'{o}nsson, F., G\'{u}stafsson, O.,
  Stef\'{a}nsson, K. and Donnelly, P. (2002). Assessing population
  differentiation and isolation from single-nucleotide polymorphism data. {\em
  Journal of the Royal Statistical Society Series B\/} {\bf 64}, 695--715.

\bibitem[Papaspiliopoulos {\em et~al.\/}(2003)Papaspiliopoulos, Roberts and
  Sk\"{o}ld]{Papaspiliopoulos/Roberts/Skold:2003}
Papaspiliopoulos, O., Roberts, G.~O. and Sk\"{o}ld, M. (2003). {Non-centred
  parameterisations for hierarchical models and data augmentation (with
  discussion)}. In: {\em {Bayesian statistics 7}\/} (eds. J.~M. Bernardo, M.~J.
  Bayarri, J.~O. Berger, A.~P. Dawid, D.~Heckerman, A.~F.~M. Smith and
  M.~West), Clarendon Press, London.

\bibitem[Patterson {\em et~al.\/}(2006)Patterson, Price and
  Reich]{Patterson:2006}
Patterson, N., Price, A.~L. and Reich, D. (2006). Population structure and
  eigenanalysis. {\em PLoS genetics\/} {\bf 2}(12), e190.

\bibitem[Pitt and Shephard(1999{\natexlab{a}})]{Pitt/Shephard:1999b}
Pitt, M.~K. and Shephard, N. (1999{\natexlab{a}}). Analytic convergence rates,
  and parameterization issues for the {G}ibbs sampler applied to state space
  models. {\em Journal of Time Series Analysis\/} {\bf 20}, 63--85.

\bibitem[Pitt and Shephard(1999{\natexlab{b}})]{Pitt/Shephard:1999}
Pitt, M.~K. and Shephard, N. (1999{\natexlab{b}}). {Filtering via simulation:
  auxiliary particle filters}. {\em Journal of the American Statistical
  Association\/} {\bf 94}, 590--599.

\bibitem[Pitt {\em et~al.\/}(2012)Pitt, {dos Santos Silva}, Giordani and
  Kohn]{Pitt:2012}
Pitt, M.~K., {dos Santos Silva}, R., Giordani, P. and Kohn, R. (2012). {On some
  properties of Markov chain Monte Carlo simulation methods based on the
  particle filter}. {\em Journal of Econometrics\/} {\bf 171}(134-151).

\bibitem[Price {\em et~al.\/}(2006)Price, Patterson, Plenge, Weinblatt, Shadick
  and Reich]{Price:2006}
Price, A.~L., Patterson, N.~J., Plenge, R.~M., Weinblatt, M.~E., Shadick, N.~A.
  and Reich, D. (2006). Principal components analysis corrects for
  stratification in genome-wide association studies. {\em Nature genetics\/}
  {\bf 38}(8), 904--909.

\bibitem[Pritchard {\em et~al.\/}(2000{\natexlab{a}})Pritchard, Stephens and
  Donnelly]{Pritchard/Stephens/Donnelly:2000}
Pritchard, J.~K., Stephens, M. and Donnelly, P. (2000{\natexlab{a}}). Inference
  of population structure using multilocus genotype data. {\em Genetics\/} {\bf
  155}, 945--959.

\bibitem[Pritchard {\em et~al.\/}(2000{\natexlab{b}})Pritchard, Stephens,
  Rosenberg and Donnelly]{Pritchard/Stephens/Rosenberg/Donnelly:2000}
Pritchard, J.~K., Stephens, M., Rosenberg, N.~A. and Donnelly, P.
  (2000{\natexlab{b}}). Association mapping in structured populations. {\em
  American Journal of Human Genetics\/} {\bf 67}, 170--181.

\bibitem[Rasmussen {\em et~al.\/}(2011)Rasmussen, Ratmann and
  Koelle]{Rasmussen:2011}
Rasmussen, D.~A., Ratmann, O. and Koelle, K. (2011). Inference for nonlinear
  epidemiological models using genealogies and time series. {\em PLoS Comput
  Biol\/} {\bf 7}(8), e1002136.

\bibitem[Roberts and Rosenthal(2001)]{Roberts/Rosenthal:2001}
Roberts, G.~O. and Rosenthal, J.~S. (2001). Optimal scaling for various
  metropolis-hastings algorithms. {\em Statistical Science\/} {\bf 16},
  351--367.

\bibitem[Rosenberg {\em et~al.\/}(2002)Rosenberg, Pritchard, Weber, Cann, Kidd,
  Zhivotovsky and Feldman]{Rosenberg:2002}
Rosenberg, N.~A., Pritchard, J.~K., Weber, J.~L., Cann, H.~M., Kidd, K.~K.,
  Zhivotovsky, L.~A. and Feldman, M.~W. (2002). Genetic structure of human
  populations. {\em Science\/} {\bf 298}, 2381--2385.

\bibitem[Sherlock {\em et~al.\/}(2013)Sherlock, Thiery and
  Roberts]{Sherlock:2013}
Sherlock, C., Thiery, A.~H. and Roberts, G.~O. (2013). {On the efficiency of
  pseudo marginal random walk Metropolis algorithms}. {\em Submitted\/} .

\bibitem[Storvik(2002)]{Storvik:2002}
Storvik, G. (2002). Particle filters for state-space models with the presence
  of unknown static parameters. {\em IEEE Transaction on Signal Processing\/}
  {\bf 50}, 281--289.

\bibitem[Wood {\em et~al.\/}(2014)Wood, {vand de Meent} and
  Mansinghka]{Wood:2014}
Wood, F., {vand de Meent}, J.~W. and Mansinghka, V. (2014). A new approach to
  probabilistic programming inference. In: {\em AISTATS\/}.

\end{thebibliography}

\end{document}